\documentclass[12pt]{article}
\usepackage{amsmath}
\usepackage{graphicx,psfrag,epsf}
\usepackage{enumerate}
\usepackage[authoryear,round,comma]{natbib}
\usepackage{url} 
\usepackage{caption}
\usepackage{amsfonts}
\usepackage{subcaption}
\usepackage{setspace}
\usepackage{JASA_manu}
\usepackage{mathtools}
\usepackage{tikz}
\usetikzlibrary{trees}

\DeclareMathOperator*{\minimize}{minimize}
\newcommand{\blind}{1}

\usepackage[figuresright]{rotating}

\begin{document}
\def\spacingset#1{\renewcommand{\baselinestretch}%
{#1}\small\normalsize} \spacingset{1}

\if1\blind
{
  \title{\bf Fair Regression for Health Care Spending}
  \author{Anna Zink\\
    Harvard University\\
    and \\
    Sherri Rose\\
    Harvard Medical School\thanks{Anna Zink is a PhD student in Health Policy (Methods), Harvard University, Cambridge, MA 02138; and Sherri Rose is an Associate Professor, Harvard Medical School, Boston, MA 02115 (e-mail: {rose@hcp.med.harvard.edu}).}}
  \maketitle
} \fi

\if0\blind
{
  \bigskip
  \bigskip
  \bigskip
  \begin{center}
    {\LARGE\bf Fair Regression for Health Care Spending}
\end{center}
  \medskip
} \fi

\bigskip
\begin{abstract}
The distribution of health care payments to insurance plans has substantial consequences for social policy. Risk adjustment formulas predict spending in health insurance markets in order to provide fair benefits and health care coverage for all enrollees, regardless of their health status. Unfortunately, current risk adjustment formulas are known to underpredict spending for specific groups of enrollees leading to undercompensated payments to health insurers. This incentivizes insurers to design their plans such that individuals in undercompensated groups will be less likely to enroll, impacting  access to health care for these  groups. To improve risk adjustment formulas for undercompensated groups, we expand on concepts from the statistics, computer science, and health economics literature to develop new fair regression methods for continuous outcomes by building fairness considerations directly into the objective function. We additionally propose a novel measure of fairness while asserting that a suite of metrics is necessary in order to evaluate risk adjustment formulas more fully. Our data application using the IBM MarketScan Research Databases and simulation studies demonstrate that these new fair regression methods may lead to massive improvements in group fairness (e.g., 98\%) with only small reductions in overall fit (e.g., 4\%). 

\end{abstract}

\vfill

\newpage

\spacingset{1.45} 

\section{Introduction}
\label{sec:intro}

Risk adjustment is a method for correcting payments to health insurers such that they reflect the cost of their enrollees relative to enrollee health. It is implemented by most federally regulated health insurance markets in the United States, including Medicare Advantage and the individual health insurance Marketplaces created by the Affordable Care Act, to prevent losses to insurers who take on sicker enrollees \citep{Pope04, McGuire14a, Kautter14}. Current risk adjustment formulas use ordinary least squares (OLS) linear regression to predict health plan payments with select demographic information and diagnosis codes from medical claims. These OLS-based formulas are then typically evaluated with overall measures of statistical fit, such as $R^2$. 

While $R^2$ is an important benchmark for evaluating global fit, it lacks information on other dimensions. As a result, risk adjustment has been criticized for not incentivizing efficient payment systems, spending, or population health management  \citep{AshEllis12, Layton17}, and for poorly estimating health costs for some groups by underpredicting their spending relative to average observed spending in the group. Underpredicting spending leads to undercompensation to the insurer, and there is evidence that insurers adjust the prescription drugs, services, and providers they cover (i.e., benefit design) to make health plans less attractive for enrollees in undercompensated groups \citep{Shepard16, Carey17, Geruso17}. Examples of  undercompensated groups include enrollees with specific medical conditions, high-cost enrollees, and partial-year enrollees \citep{VanKleef13, Montz16, Geissler17}. Recent research has also shown that health plan insurers have the ability to identify undercompensated groups \citep{Jacobs15, Geruso17, Rose17, VanKleef18}.

What constitutes a fair or unfair algorithm depends heavily on the context.  These fairness concepts and methods have been largely developed in the computer science literature \citep{Chouldechova18}. 
We will consider risk adjustment formulas unfair if they underpredict spending for a prespecified group of enrollees, which then incentivizes differential treatment for the group via benefit design due to this undercompensation. For example, if average observed spending for individuals with mental health and substance use disorders (MHSUD) is $\$$10,000, but  average predicted spending  in this group is $\$$8,000, the risk adjustment formula may be unfair for the MHSUD group by `substantially' underpredicting their spending. We define formal  metrics for evaluating fairness in risk adjustment formulas using group residual errors in the next section.

Methods for addressing fairness are often separated into three categories based on the point in the learning process at which fairness is addressed: the preprocessing, fitting, or postprocessing phase. 
If the data are inherently biased, then preprocessing techniques are a possible solution. These methods create fair datasets by transforming or changing the data so that it is no longer biased \citep{Kamiran12, Zliobaite11, Zemel13, Calmon17, Johndrow17}. It has been shown that current spending patterns among various group may be undesirable due to the plan benefit system, and by using observed spending data, we reinforce these unfair spending patterns. 
A recent study  explored this concept by transferring funds to undercompensated groups in the raw data in order to promote more ideal spending patterns  \citep{Bergquist19}.

One of the most common fitting phase approaches in risk adjustment attempts to fix group undercompensation by adding new variables representative of the groups in the risk adjustment formula \citep{VanKleef13}. While this is a straightforward idea, it can be problematic if those variables are unavailable, incentivize over- or underutilization of health services, or the risk adjustment formula does not recognize the improvement \citep{RoseMcguire19}.  Fitting techniques in fairness include separate formulas for protected classes as well as  fairness penalty terms or  constraints \citep{Kamishima12,Berk17b, Zafar17b, Zafar17,Bechavod17, Dwork17}. We see intersections of these areas  in the risk adjustment literature with separate formulas for enrollees with MHSUD \citep{Shrestha17,VanKleef18c} and constrained regression to reduce undercompensation for specific groups \citep{VanKleef17b}.   Notably, separate risk adjustment formulas are already used in practice for infants and adults due to   known differences in spending patterns. 
Nonparametric statistical machine learning methods to enhance  estimation accuracy in risk adjustment have also been explored for the fitting stage \citep{Rose16, Shrestha17, Park18}, but none of these tools  are currently deployed in the U.S. health care system.

Postprocessing techniques modify the results  after fitting by, for example, creating specific classification thresholds for different groups \citep{Bansal14,Hardt16,Kleinberg18,Mhamdi18}. These methods separate fit  from fairness objectives and allow  use of the same prediction function for multiple fairness objectives. Reinsurance, paying insurers for a portion of the costs of high-cost enrollees,  can be considered  postprocessing for risk adjustment in that it reduces undercompensation for high-risk enrollees \citep{rabook}. 

In this paper, we focus on the fitting phase and expand on concepts from  statistics, computer science, and health economics, proposing new estimation methods and measures to improve risk adjustment formulas for undercompensated groups. 
We develop two new fair regression estimators for continuous outcomes that reduce residual errors for an undercompensated group by  building fairness considerations directly into the objective function. We also extend a definition of fairness from the computer science and statistics literature for the risk adjustment setting while additionally considering existing measures. 

Our  application features the IBM MarketScan Research Databases. This set of databases contains enrollee-level claims, demographic information, and health plan spending for a sample of individuals (and their dependents) insured by private health plans and large employers across the country. In 2014, the IBM MarketScan Research Databases were used by the federal government to develop the risk adjustment formula for the individual health insurance Marketplaces. Thus, this data source is particularly policy relevant.  The undercompensated group we focus on for this data application is enrollees with MHSUD. We select this group for two major  reasons. First, individuals with MHSUD are known to have substantially undercompensated payments in current risk adjustment formulas \citep{Montz16}. Second, about 20\% of people in the United States have  MHSUD, thus it is a  priority area for  policy change.  Although the data are representative of only a subset of the U.S. health insurance market, our methods are appropriate for other markets and different application settings with continuous outcomes. The methods and metrics we present are compared in this data analysis as well as simulation studies.

\section{Statistical Framework}
This section describes our approach to fair regression. It involves a suite of fairness measures for evaluating new and existing regression tools in an effort to improve risk adjustment formulas for undercompensated groups. A typical algorithmic fairness problem has an outcome $Y$ and input vector $\boldsymbol{X}$ that includes a protected group $A 
\subset \boldsymbol{X}$. The goal is to create an estimator for the function $f(\boldsymbol{X})=Y$ that maps $\boldsymbol{X}$ to   $Y$, while aiming to ensure that the function is fair for protected group $A$. Although our main goal is to understand whether  estimation methods beyond OLS, including those we newly propose, improve fairness for risk adjustment, we also wish to focus on interpretability for stakeholders, such as government agencies, insurers,  providers, and enrollees. Therefore, constrained and penalized regressions were natural choices to enforce fairness in risk adjustment for undercompensated groups. 

\subsection{Measures}\label{sec:measures}

The most commonly used measures of fairness  are based on the notion of group fairness, striving for similarity in predicted outcomes or errors for groups. 
Let $g$ be the set containing all $n_g$ enrollees with MHSUD (i.e., the undercompensated group), indexed by $i$. 
The complement group, all  $n_{c}$ enrollees without MHSUD, is denoted by $g^c$  and  indexed by $j$. Overall sample size, $N=n_g+n_{c}$, is indexed by $k$. Group undercompensation is a  result of large average group residuals in the risk adjustment formula.  We  define fairness as a function of these residual errors given that many undercompensated groups have  substantially higher average health care costs. Thus, enforcing similar predicted outcomes $\hat{Y}$ between $g$ and $g^c$ would be unfair to both. In this subsection, we present three relevant existing measures of group fairness, a new extension of fair covariance modified for group fairness with continuous outcomes, and $R^2$ as a metric of overall global fit.

\textit{Mean Residual Difference.}  Comparing mean residual errors between a group $g$ and its complement $g^c$ aims to assess fairness by evaluating whether this difference is close to zero \citep{Calders13}: 
	$1/n_g\sum_{i \in g} (\hat{Y}_i-Y_i) - 1/n_{c}\sum_{j\in g^c} (\hat{Y}_j-Y_j).$
To date, this metric has not been applied in risk adjustment.

\textit{Net Compensation.} Net compensation is a  related measure from the health economics literature on the same scale as the mean residual difference \citep{Layton17}. However, it does not contain a term for the mean residual in the complement group:
$1/n_g \sum_{i \in g} (\hat{Y}_i-Y_i).$ Therefore, this measure focuses on a reduction in the residuals for $g$ rather than similarity in residuals between the  groups. A parallel net compensation measure can be calculated for $g^c$. 

We highlight that we intentionally take the difference $\hat{Y}_i-Y_i$ rather than $Y_i-\hat{Y}_i$ so that undercompensation for those in $g$ aligns with a negative value of net compensation, in line with previous literature \citep[e.g.,][]{Bergquist19}. This is reflected in the mean residual difference definition above as well. We do not maintain this ordering for the corresponding estimators in Section~\ref{est}  as we wish to penalize large undercompensation in net compensation penalized  regression by \textit{adding} to the squared error and the squared  term for mean residual difference penalized regression negates the ordering distinction.

 \textit{Predictive Ratios.} Predictive ratios are commonly used to quantify the underpayment for specific groups in risk adjustment \citep{Pope04}: $ \sum_{i \in g}\hat{Y}_i/\sum_{i \in g} Y_i.$ Whereas net compensation provides the absolute magnitude of the loss in dollars, predictive ratios provide the relative size of the loss. Predictive ratios can also be created for $g^c$. 

\textit{Fair Covariance.} Other fairness work creates a measure based on the idea that to be fair, the predicted outcome (or  residual error) and protected class must be independent. Using the covariance between the predicted outcome (or residual error) and the protected class as a proxy for independence, that work establishes a fairness measure \citep{Zafar17b,Zafar17}. Because this prior metric assumes outcomes are classified into discrete categories, we extend the definition  to define a new measure of fair covariance for residual errors with continuous $Y$. 
Our measure is  given by: $Cov(A, Y-\hat{Y})$, where $A\in\{0,1\}$ is the random variable indicating membership in $g$. This measure is bounded by the covariance of the undercompensated group and the OLS residual, which we refer to as $c^*$. Our fair covariance measure  allows one to see the empirical signal for systematic undercompensation through residual covariance and it can also be scaled by $c^*$ such that it is bounded between 0 and 1. 

 \textit{Global Fit.} In addition to fairness measures, we also evaluate overall fit with the traditional measure used in risk adjustment, which is 
 $R^2$:
$1-\sum_{k}(Y_k-\hat{Y}_k)^2/\sum_{k}(Y_k-\bar{Y}_k)^2$,
\noindent where we recall that  $k$ indexes the overall sample. Given  current  policymaker prioritization of global metrics, it is important to compare estimators with both group and overall fit measures to understand the impact on global  fit when seeking fairness for undercompensated groups. 

The measures we consider above assume that the data include unbiased $Y$, which may not be the case in practice. Additionally, fairness is frequently assessed for one or two groups, as we also do here. In reality, we are often concerned about fairness for many groups. This requires the ability to {define all meaningful groups}, which is not always an objective task.  
There are also tradeoffs involved in selecting a fairness metric, and ensuring fairness based on one definition does not necessarily guarantee a satisfying solution with respect to other fairness measures or  overall fit \citep{Kleinberg16, Chouldechova17, Berk17}.  We return to these issues in our discussion. 
In Web Appendix A, we present a new extension of  a  fairness measure for comparing individual residual errors rather than mean residual errors. This \textit{Group Residual Difference} metric is not practical to implement at scale in risk adjustment, thus we do not deploy it here, but could be useful for settings where $N$ is smaller.

\subsection{Estimation Methods} 
\label{est}

We present five methods that incorporate a fairness objective with a constraint or penalty to improve risk adjustment formulas for undercompensated groups. Two of these methods, covariance constrained regression and net compensation penalized regression, are new contributions, and all five methods will also be compared to the OLS estimator. 
We have a continuous spending outcome $Y$, a vector of binary health variables $\boldsymbol{H}=(H_1,\ldots, H_T)$,  an input vector $\boldsymbol{X}=\{\text{female}, \text{age}, \boldsymbol{H}\}$, and a coefficient vector $\boldsymbol{\theta}$  indexed by $p$.  For OLS, we aim to solve the following regression problem:
\begin{align}
&\minimize_{\theta} \left\{\sum_{k} \bigg(Y_k- \sum_{p}\theta_p X_{kp}\bigg)^2 \right\}.
\label{olsform}
\end{align}

\textit{Average Constrained Regression.}  A previously proposed constrained regression method for risk adjustment requires that the estimated average spending for the undercompensated group is equal to the average spending, which means that  net compensation for the undercompensated group is zero \citep{VanKleef17b}. This is achieved by including a constraint:
$\minimize_{\theta} \{\sum_{k} (Y_k- \sum_{p}\theta_p X_{kp})^2 \}$, subject to  $1/n_g\sum_{i \in g} Y_i = 1/n_g\sum_{i \in g} (\sum_{p}\theta_p X_{ip}).$ The given constraint has been applied in the risk adjustment literature to reduce undercompensation for select groups \citep{VanKleef17b,Bergquist19}.

\textit{Weighted Average Constrained Regression.} The next existing method relaxes the previous constraint, allowing the estimated spending to be a weighted average of  the average spending of the undercompensated group and the estimated spending under unconstrained OLS:
 $\minimize_{\theta} \{\sum_{k} (Y_k- \sum_{p}\theta_p X_{kp})^2 \}$, 
subject to $1/n_g\sum_{i \in g} (\sum_{p}\theta_p X_{ip} ) = (1-\alpha)/n_g\sum_{i \in g} Y_i +  \alpha/n_g\sum_{i \in g} (\sum_{p}\theta^{OLS}_p X_{ip}),$
 where $\boldsymbol{\theta}^{OLS}$ is the coefficient vector from the  OLS given in formula~(\ref{olsform}). The hyperparameter $\alpha \in [0,1]$ is a weighting factor. When $\alpha=0$, this method is  equivalent to average constrained regression, and when  $\alpha=1$ it is equivalent to  OLS. Weighted average constrained regression has been shown to reduce undercompensation for select groups in the Netherlands risk adjustment formula \citep{VanKleef17b}.

\textit{Covariance Constrained Regression.} The class of covariance methods we consider impose a constraint on the residual by requiring that the covariance between the residual and the protected class is close to zero \citep{Zafar17b,Zafar17}. We extend these techniques to propose a new method for our risk adjustment setting where we have a continuous residual, which has not been previously explored. In order to solve the optimization problem, we convert it into a convex problem. 
We simplify the covariance as follows: 
\begin{align*}
Cov(A,Y-\boldsymbol{\theta}\boldsymbol{ X}) &= E[\{A-E(A)\}\{Y-\boldsymbol{\theta X}-E(Y-\boldsymbol{\theta X})\}]\\
& = E[\{A-E(A)\}(Y-\boldsymbol{\theta X})] \\
&\approx \frac{1}{N} \sum_k \left[\{A_k-P(A=1)\}\bigg(Y_k-\sum_{p}\theta_p X_{kp}\bigg)\right] \\
&\approx \frac{1}{N} \Bigg[\{1-P(A=1)\} \sum_{i\in g} \bigg(Y_i-\sum_{p}\theta_p X_{ip}\bigg) \\
& \hspace{66pt}-P(A=1)\sum_{j \in g^c} \bigg(Y_j-\sum_{p}\theta_p X_{jp}\bigg)  \Bigg].
\end{align*}
\noindent Now that we have the covariance in the form of a convex problem, we can define what we need to solve:
$\minimize_{\theta} \{\sum_{k} (Y_k- \sum_{p}\theta_p X_{kp})^2 \}$, 
 subject to $\{1-P(A=1)\} \sum_{i\in g} (Y_i-\sum_{p}\theta_p X_{ip})  - P(A=1)\sum_{j \in g^c} (Y_j-\sum_{p}\theta_p X_{jp}) <c$ and 
$\{1-P(A=1)\} \sum_{i\in g} (Y_i-\sum_{p}\theta_p X_{ip})  - P(A=1)\sum_{j \in g^c} (Y_j-\sum_{p}\theta_p X_{jp})\geq -c$.
Parallel to the literature for discrete categories \citep{Zafar17}, we  set $c=m\times c^*$, where $m$ is a multiplicative factor $m \in [0,1]$ and $c^*$ is the covariance of the undercompensated group and the OLS residual. The upper bound for $c$ occurs at $m=1$, which is $c^*$. 

As we are primarily concerned with the residual of the undercompensated group being too large, we choose to instead bound the covariance on one side in our implementation of this method.  In other words, we constrain the covariance to be less than some percentage of the OLS covariance (as defined by the hyperparameter $m$).  A one-sided constraint also yields faster optimization. The updated optimization problem is: 
$\minimize_{\theta} \{\sum_{k} (Y_k- \sum_{p}\theta_p X_{kp})^2 \}$, 
subject to $\{1-P(A=1)\} \sum_{i\in g} (Y_i-\sum_{p}\theta_p X_{ip})  - P(A=1)\sum_{j \in g^c} (Y_j-\sum_{p}\theta_p X_{jp}) <c$.

\textit{Mean Residual Difference Penalized Regression.} The relationship between penalized and constrained regressions is well recognized in  statistics \citep{statlearn}, and one could equivalently reformulate the above constraints as penalties. Penalized regression has also been explored in the fairness literature. \cite{Calders13}  consider constrained formulations of their approaches, but propose the flexibility of penalization as an alternative due to the possibility of degenerate solutions with a high number of constraints.
  In their mean residual difference regression technique, one penalizes  with large mean residual differences between the undercompensated group and the complement group. The coefficients minimize: 
$ \sum_{k} (Y_k- \sum_{p}\theta_p X_{kp})^2   +\lambda \{1/n_g\sum_{i \in g} (Y_i-\sum_{p}\theta_p X_{ip}) - 1/n_c\sum_{j\in g^c} (Y_j-\sum_{p}\theta_p X_{jp})\}^2$, 
where  hyperparameter $\lambda$ can be user-specified or chosen via cross-validation, and its magnitude will be on the same scale as $Y$. 

\textit{Net Compensation Penalized Regression.} In our second new method, rather than imposing a constraint, we also formulate a penalized regression. Our regression involves the inclusion of a custom net compensation penalty term  in the minimization problem:  
$\sum_{k} (Y_k- \sum_{p}\theta_p X_{kp})^2  + \lambda \{1/n_g\sum_{i \in g} (Y_i-\sum_{p}\theta_p X_{ip})\}$.
 This penalty punishes estimators where the net compensation, or difference between the average spending and predicted spending for the undercompensated group, is large. We can alternatively present our new method as a constraint:  $\minimize_{\theta} \{\sum_{k} (Y_k- \sum_{p}\theta_p X_{kp})^2 \}$, subject to  $1/n_g\sum_{i \in g} (Y_i-\sum_{p}\theta_p X_{ip}) \leq z$,
where the hyperparameter $z$ is positive and has a one-to-one correspondence with, but is not equal to, $\lambda$ when the constraint is binding. We choose to primarily implement this method as a penalized regression to explore differences in performance with the mean residual difference penalized regression for the same values of $\lambda$. However, simulation studies in Web Appendix B of the Supplementary Material examine  the performance of the constrained formulation.

\subsection{Computational Implementation} 
These six methods were evaluated  to assess both overall fit and fairness goals with 5-fold cross-validation in our data analysis and simulations using the suite of five measures defined in Section~\ref{sec:measures}.  
 OLS  was implemented in the \texttt{R} programming language with the \texttt{lm()} function. All other estimators were optimized using the \texttt{CVXR} package. This package uses disciplined convex programming to solve optimization problems and allows users to  specify novel constraints and penalties \citep{cvxr}.

\section{Health Care Spending Application}

We selected a random sample of 100,000 enrollees from the IBM MarketScan Research Databases. Age, sex, and diagnosed health conditions, all from the year 2015, were used to predict total annual expenditures in 2016. 
Diagnosed health conditions took the form of the established Hierarchical Condition Category (HCC) variables created for risk adjustment. HCCs were  developed by the Department of Health and Human Services to group  a selection of International Classification of Disease and Related Health Problems (ICD) codes into  indicators for various health conditions \citep{Pope04, Kautter14}. 
We considered the 79 HCC variables currently used in Medicare Advantage risk adjustment formulas and retained the 62 HCCs that had at least 30 enrollees with the  condition. See Web Appendix C for a list of the 62 HCCs included in the regression formulas. Our sample of enrollees was 52\% female and between the ages of 21 and 63, with median age 45. Mean and median annual expenditures per enrollee were \$6,651 and \$1,511, respectively.

We defined enrollees with MHSUD, our protected group $A$, using Clinical Classification Software (CCS) categories. This classification system maps each MHSUD-related ICD code to a CCS category, unlike the HCCs, which only map a subset of MHSUD-related ICD codes. Based on CCS categories,  13.8\% of the sample had a diagnosis code for MHSUD compared to 2.6\% had we used  HCCs. We note that we do not capture enrollees with MHSUD who do not have an ICD code for their condition(s). The mean annual expenditures for MHSUD enrollees in our sample were \$11,520 versus \$5,880 for enrollees without MHSUD (and \$3,744 versus \$1,274 for median annual expeditures).

\begin{figure}

\includegraphics{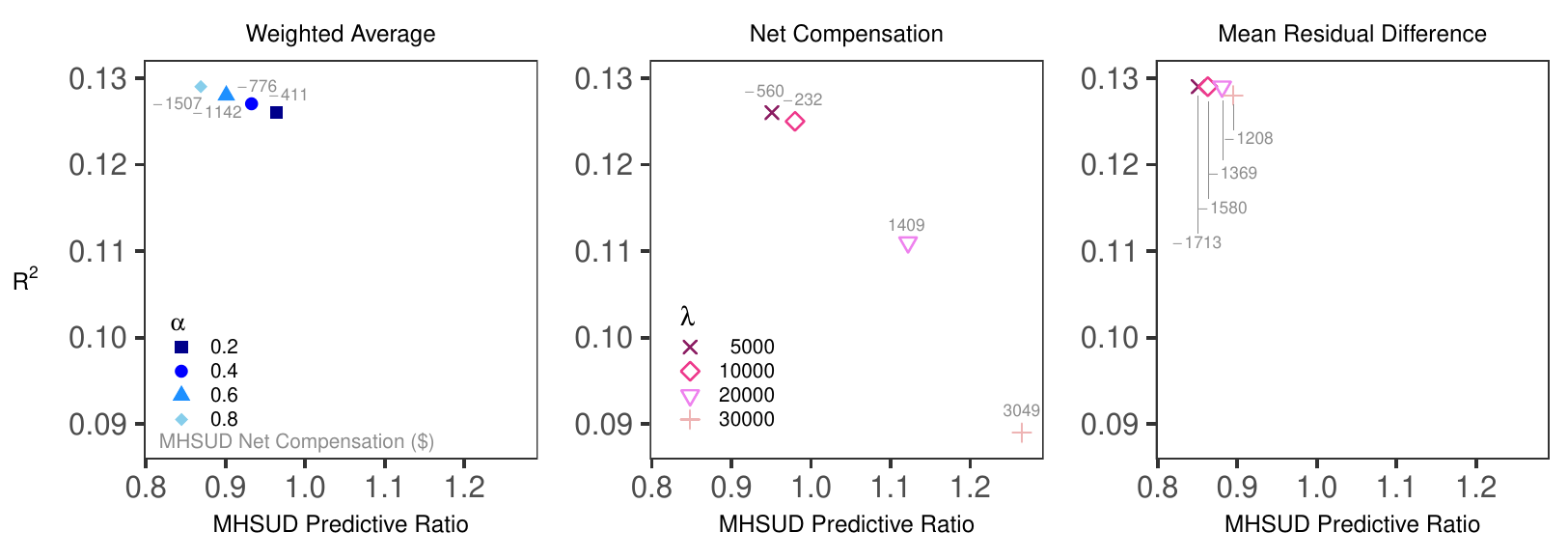}
\caption{\textbf{Global Fit versus Group Fairness.} Variation in cross-validated performance by hyperparameter is plotted for three estimators. Predictive ratios for mental health and substance use disorders (MHSUD) are contrasted with overall $R^2$ fit. Results for all hyperparameters  in the covariance constrained regression, $m \in \{0.2,0.4,0.6,0.8\}$, were extremely similar and thus omitted.}
\label{plot:tuning}
\end{figure}

We compared each method to determine which estimators were best at reducing undercompensation for enrollees with MHSUD, and at what cost to overall statistical fit. In Table~\ref{tab:commands}, we report the top estimators with respect to fairness for each of the six methods,  having selected the hyperparameter value  that optimizes the fairness measures (for those that have these parameters). Hyperparameter values were user-specified from the range of plausible values. For example, in the covariance constrained regression, $m$ can range from 0 to 1, and we considered $m \in \{0.2,0.4,0.6,0.8\}$. Comparisons of global fit versus group fairness for the three methods with variation in performance by hyperparameter can be found in Figure~\ref{plot:tuning}. 
 
\begin{table}
\small
  \caption{Performance of Constrained and Penalized Regression Methods}
  \begin{tabular}{rcccrrrr}
     & & \multicolumn{2}{c}{Predictive} & \multicolumn{2}{c}{Net} & \multicolumn{1}{c}{Mean}  & \multicolumn{1}{c}{}\\ 
     & & \multicolumn{2}{c}{Ratio} & \multicolumn{2}{c}{Compensation} &  \multicolumn{1}{c}{Residual} & \multicolumn{1}{c}{Fair}\\ 
  \textbf{Method}    &  $R^2$   & $g$ & $g^c$ & \multicolumn{1}{c}{$g$} & \multicolumn{1}{c}{$g^c$} & Difference & Covariance \\    
\noalign{\medskip}\hline\noalign{\medskip}
    Average  & 12.4\% &   0.996 & 1.001 &-\$46 & \$4 & -\$50 & 6\\ 
    \\
    Covariance & 12.4 &  0.996 & 1.001 & -46 & 4 &  -50 & 6\\
    \\
    Net Compensation$^\dagger$  & 12.5 &   0.980 & 1.006 & -232 & 34 &  -266 & 31\\
 \\
         Weighted Average$^\mp$    & 12.6 &  0.964 & 1.011 & -411 & 62 &  -473 & 56\\ 
          \\
    Mean Residual Difference$^\oplus$  & 12.8 &  0.895 & 1.032 &-1208 & 188 & -1396 & 164\\
    \\
        OLS & 12.9 &  0.837 & 1.050 & -1872 & 293 & -2165 & 256 \\ 
\noalign{\medskip}\hline
  \end{tabular}
    \label{tab:commands}
 
   \vspace{22pt}
 
   \footnotesize{$^\dagger \lambda=10000$, $^\mp \alpha=0.2$, $^\oplus \lambda=30000$\\
   \textit{Note:} Measures calculated based on cross-validated predicted values and sorted on net compensation. Best performing hyperparameters  for each estimator (with respect to fairness measures) are displayed. Performance for covariance method was the same for all $m$. $g^c$ is the complement of g.}

\end{table}

 OLS  had a cross-validated $R^2$ measure of 12.9\%, a predictive ratio of 0.837 for individuals with MHSUD, and underestimated average MHSUD spending by -\$1,872, with a mean residual difference of -\$2,165. The fair covariance measure was 256.  Average spending for enrollees without MHSUD was overestimated by \$293 with a predictive ratio of 1.050. OLS had the worst performance along all fairness metrics while producing an $R^2$ only trivially higher than the competing methods.

We found the best improvement in fairness for MHSUD using the existing average constrained regression and our new covariance constrained regression. These two methods had similar, although not identical performance, and reduced the average undercompensation for enrollees with MHSUD to -\$46 (versus -\$1,872 in the  OLS), a relative  improvement of 98\%. They also increased the predictive ratio from 0.837 to 0.996. Enrollees without MHSUD were overestimated by only \$4 and had a predictive ratio of 1.001. Both methods reduced the fair covariance measure from 256 to 6. Unsurprisingly, these two estimators were also the  worst performers on overall fit as measured by $R^2$, although it was a loss of only 4\%, from 12.9\% to 12.4\%. This small 0.5 percentage point loss in $R^2$ may be tolerable to policymakers.

Recall that the weighted average constrained regression is a compromise estimator between the OLS and average constrained regression.
As $\alpha$ approached one in the first panel of Figure~\ref{plot:tuning}, the metrics more closely resembled the OLS results. As $\alpha$ approached zero we saw values closer to the average constrained regression results, although $\alpha=0.2$ was not only dominated by the average constrained  and covariance constrained regressions, but also the net compensation penalized regression with $\lambda=10000$.

The remaining two methods were  regressions with customized penalty terms to punish unfair estimates. Our proposed net compensation penalized regression  varied substantially by  hyperparameter (see second panel in Figure~\ref{plot:tuning}), although was the third best performer overall when $\lambda=$10000. Large $\lambda$ values yielded extremely poor performance on both overall fit and fairness. At $\lambda=20000$, $R^2$ dropped by 12\% to 11.9\%, and when $\lambda$ increased to $30000$, $R^2$ dropped to 9\%, a relative reduction of 29\%. These two $\lambda$ values led to a large \textit{over}compensation for enrollees with MHSUD. The covariance was also negative, indicating that the residual value for MHSUD was systematically too high. The mean residual difference penalized regression was less sensitive to hyperparameters compared to the net compensation penalized regression (see third panel in Figure~\ref{plot:tuning}). The best performance for mean residual difference penalized regression was at $\lambda=30000$; it improved on the MHSUD predictive ratio for OLS by 7\% (from 0.837 to 0.895) with an $R^2$ loss of less than 1\%. However, the best performing net compensation penalized regression had an 81\% improvement over the best performing mean residual difference penalized regression when comparing MHSUD net compensation, as well as large improvements in predictive ratios (0.895 versus 0.980) and fair covariance (164 versus 31).

\begin{figure}
\begin{center}

\includegraphics[height=6in]{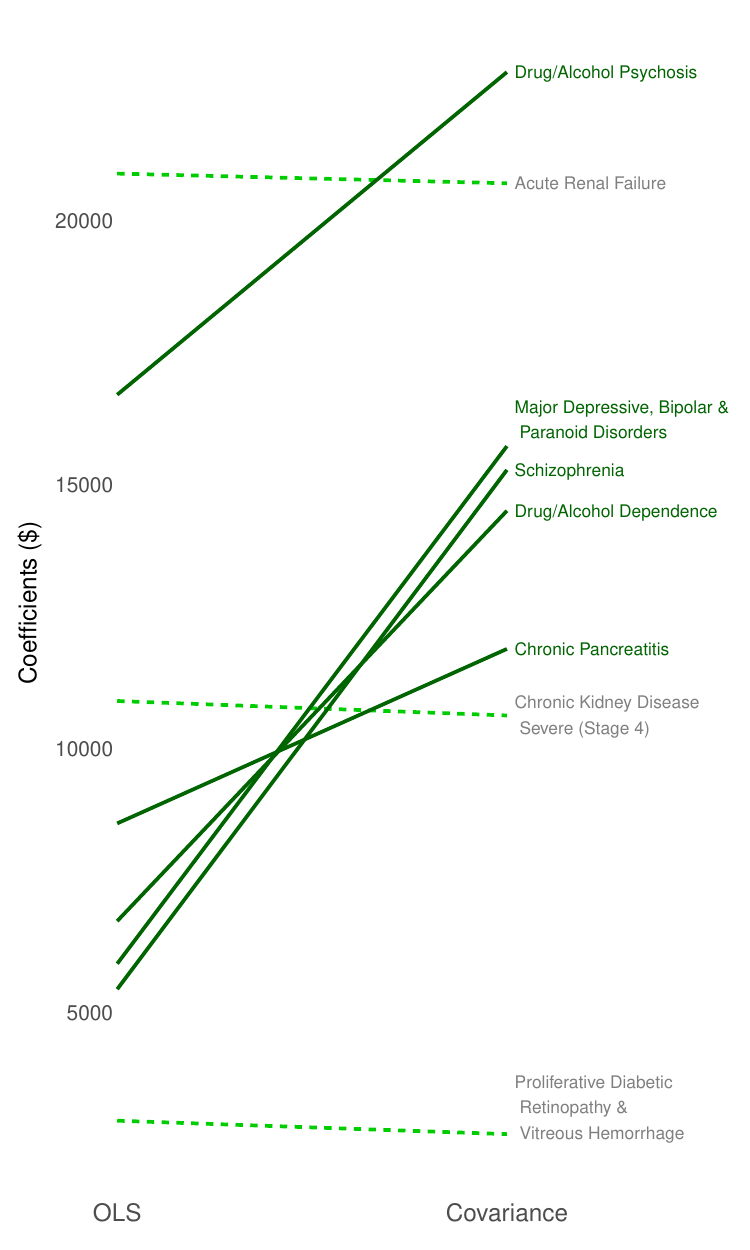}
\end{center}
\caption{\textbf{Largest Coefficient Changes.} Increases in  coefficient values from the OLS to covariance constrained regression are represented by  solid lines with decreases in dashed  lines. Largest five increases and largest five decreases were considered; ``Chronic Kidney Disease, Severe (Stage 4)'' and ``Severe Hematological Disorders'' (both decreases) were suppressed due to  large magnitudes while  having small relative percentage changes of $<$1\%}

  \label{plot:coef}
\end{figure}

We also examined the HCC variable coefficients for the best performing estimators, the average constrained  and  covariance constrained regressions, in comparison to OLS. Risk adjustment coefficients communicate incentives to insurers and providers related to prevention and care. For example, coefficients that do not reflect costs can impact an insurer's incentives in creating their plan offerings. Coefficients for the average constrained  and covariance constrained regressions were nearly identical when rounded to the nearest whole dollar, thus we display OLS versus covariance constrained regression in Figure~\ref{plot:coef}. We considered the largest five increases and largest five decreases from OLS to covariance constrained regression, and observed sizable increases in the estimated coefficients associated with MHSUD. The largest relative increase was 180\% for ``Schizophrenia.'' Relative decreases were much smaller.

\section{Simulation Study}

A set of simulation scenarios was developed to explore how these regression methods perform in other settings. We generated a population of 100,000 observations with two continuous outcomes $Y_1$ and $Y_2$ that were each a function of covariates in $\boldsymbol{X}=(X_1, X_2, ..., X_9)$ and two distinct yet partially overlapping protected classes ($A_1$ and $A_2$) that depended on variables in $\boldsymbol{X}$. 
Scenario 1 considered a complex functional form for $Y_1$ and regression estimators that were misspecified, including omitted $\boldsymbol{X}$ variables. Scenario 2 examined a less complex functional form in $Y_2$ and regression estimators that were misspecified, including additional noise variables but no omitted  $\boldsymbol{X}$ variables. A third scenario is discussed in Web Appendix B of the Supplementary Material, along with complete details for the simulated population and first two scenarios. For each scenario, we drew 500 samples of $N=$1,000 and $N=$10,000 observations from the simulated population of 100,000 observations. As in the data analysis, hyperparameter values were user-specified from the range of plausible values.

\begin{figure}
\includegraphics{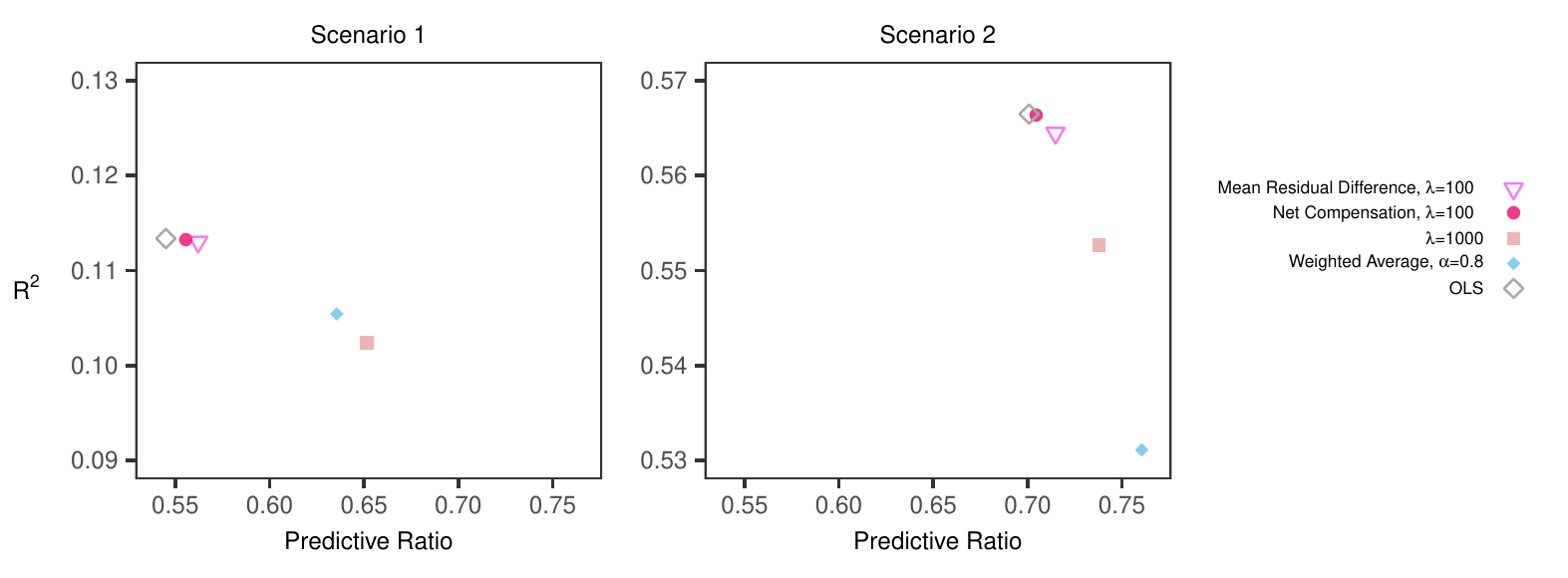}
\caption{\textbf{Simulation Results.} The plot includes OLS and estimation methods that improved fairness measures with a relative cross-validated $R^2$ loss  $\leq 10$\% for $N=10,000$. Predictive ratios for protected class $A_1$ are contrasted with overall $R^2$ fit. }
\label{plot:sim}
\end{figure}

Selected results  are presented in Figure~\ref{plot:sim}, which includes  OLS and those methods that improved fairness measures for protected class $A_1$ with a relative $R^2$ loss  $\leq 10$\%. Notably, average constrained  and covariance constrained regression, the tied top  estimators in our data analysis, do not appear.  This was common across settings;  average constrained  and covariance constrained regression often struggled with functional form misspecification. However, net compensation penalized regression, which performed well in our data analysis, also performed well in the simulations with respect to achieving metric balance between global fit decreases and group fit increases. Additional results are available in Web Appendix B of the Supplementary Material.

\section{Discussion}

We proposed new fair regression methods aiming  to  improve risk adjustment for undercompensated groups and asserted that a broader set of metrics  is needed. As expected, there was no single method that performed the best across all the measures. 
One of our  newly proposed techniques, net compensation penalized regression, had strong performance with respect to fairness and global fit  in \textit{both} the data analysis and simulations. Selecting the `best' method relies on subjective decisions regarding how to balance group fairness versus overall fit tradeoffs. Improvements in fairness resulted in subsequent decreases in $R^2$. However, for many estimators, particularly in our data analysis, improvements in fairness were larger than the subsequent decreases in overall fit. This suggests that if we allow for a slight drop in overall fit, we could greatly increase compensation for {MHSUD}. Policymakers need to consider whether they are willing to sacrifice small reductions in global fit for large improvements in fairness. 

We used a sample of enrollees in our demonstration. At scale in a  policy implementation, data from millions of enrollees would be used to estimate health spending. Solutions to group undercompensation must be scalable, and current software may or may not yet be capable of handling the sample sizes required. We tested the \texttt{CVXR} optimization package on larger samples and found that it was able to find solutions on a sample of 1,000,000 observations over the span of 3 days (versus 7 hours for the 100,000 enrollee sample). While the optimization results were not within the ideal optimal threshold, they still converged and the results were similar to those presented in this paper, which is promising. Future work includes additional studies regarding scalability. In our analyses, we also considered a selection of user-specified hyperparameter values. A more thorough approach, with possibly improved results, would explore the hyperparameter space in an automated way to select values that optimize over joint fairness and fit objectives. As a general guideline, we found that $\lambda=N/10$ yielded reasonable metric balance for our newly proposed net compensation penalized regression.

We focused on one group that risk adjustment is known to disadvantage, but it is important to extend such strategies to multiple groups. Improvements for one group could result in subsequent undercompensation for other groups, and balancing fairness across an increasing number of groups is an as yet unsolved problem in risk adjustment. Our simulations examined two protected classes, and we found that improving fairness for one group did not generally help or harm the second group. Earlier research developing methods for the preprocessing phase  found that reducing undercompensation for enrollees with MHSUD improved fairness measures for other groups, including enrollees with multiple chronic conditions but without MHSUD. Among the groups included in their comparisons, only enrollees with heart disease had slight reductions in fairness \citep{Bergquist19}. But even the act of defining the groups poses a problem, as this can be subjective, potentially favoring larger groups with well-funded advocacy organizations. Undercompensation could be undetected in many other lesser-known groups. However, we can only measure undercompensation for groups that are identified by available data, and socioeconomic information, such as poverty and housing, are not available at the individual level for risk adjustment \citep{Ellis17}.

Broadly, data-driven decisions have come under scrutiny for perpetuating human biases and disparities, which  certainly exists in risk adjustment. Arguments for a more comprehensive view of research results is increasing among scientific researchers today \citep{nyt,nature}. Recent work argues that evaluating methods from a purely statistical standpoint can lead to negative consequences, and that policy aims should be better incorporated into our research \citep{Davies18}.  Our article follows in this spirit, and we presented additional estimators and comparisons  across multiple measures for the numerous (sometimes competing) goals of risk adjustment. 
While we worked  within the specific context of risk adjustment, the fairness methods and measures discussed here have implications for other settings with continuous outcomes, which have been understudied relative to binary outcomes.

\section*{Supplementary Materials}
Web Appendix A presents the \textit{Group Residual Difference}, a new extension of  a  fairness measure for comparing individual residual errors that is not currently practical to implement at scale in risk adjustment. Web Appendix B, referenced in Sections 2 and 4, describes the simulation study details and results.  Web Appendix C is a list of HCCs included in the data analysis regression formulas from Section 3. The IBM MarketScan Research Databases used in Section 3 are not available for public dissemination as they contain protected patient information and we were granted access via a restricted  data use agreement. Instead, we generated a simulated version of the analysis data that preserves important relationships while protecting the original content, as described in Web Appendix D. We provide this simulated analysis data with code as well as code to reproduce the simulation study from Section 4 with this submission and online: \url{https://github.com/zinka88/Fair-Regression}.

\section*{Acknowledgments}

This work was supported by the National Institutes of Health through an NIH Director's New Innovator Award DP2-MD012722. The authors thank the Health Policy Data Science Lab, Thomas G. McGuire, Berk Ustun,  Jos\'{e} Zubizarreta, and the anonymous reviewers for helpful comments.

\pagebreak

\section*{Web Appendix A: Group Residual Difference}
\textit{Group Residual Difference.} In the fairness literature, one  definition for continuous outcomes is that persons with similar $Y$ should have similar predicted outcomes $\hat{Y}$ regardless of their protected class \citep{Berk17b}. 
This relies on a user-defined distance function $d$, such as $|Y_i-Y_j|$, to ensure that people who are `close'  have similar outcomes. We  extend this definition for risk adjustment by comparing residuals rather than predicted outcomes for the two groups: 
$[{1}/(n_g n_{c}) \sum_{i\in g, j \in g^c} d(Y_i, Y_j)\{Y_i-\hat{Y}_i - (Y_j-\hat{Y}_j)\}]^2.$
We refer to this new measure as the group residual difference. However, this measure is not practical to implement  at scale in risk adjustment, which often involves millions of enrollees. The group residual difference requires comparing the residual of every enrollee in the undercompensated group to every other enrollee in the complement group. This scaling issue was also noted in the earlier  work our metric extends upon \citep{Berk17b}. Our group residual difference metric can be useful in  settings where $N$ is smaller.

\pagebreak
\section*{Web Appendix B: Simulation Study Details}
As described in Section 4 of the main text, our simulation study population of 100,000 observations considered covariates $\boldsymbol{X}=(X_1, X_2, ..., X_9)$, two protected class indicator variables ($A_1$ and $A_2$), and two continuous outcome variables ($Y_1$ and $Y_2$). $X_1$ was generated from a Normal distribution with mean 70 and standard deviation 15.  Both $X_2$ and $X_3$ had Poisson distributions, with $\lambda$ values of 10 and 35, respectively. The last six covariates ($\boldsymbol{X}_{4:9}$) were drawn from Bernoulli distributions with probabilities 0.5, 0.1, 0.05, 0.8, 0.03, and 0.2. $A_1$ and $A_2$ were also drawn from Bernoulli distributions, but  depended on other generated variables in the population: 
\begin{eqnarray*}
\centering
A_1 &\sim&  \text{Bernoulli}(X_4\times X_9/2+.01) \\
A_2 &\sim& \text{Bernoulli}(X_4^2/3+.05). 
\end{eqnarray*}
They had prevalence rates of $6\%$ and $22\%$, respectively, with  2.1\% overlap.  Both outcomes, $Y_1$ and $Y_2$, depended on variables in  $\boldsymbol{X}$ as well as $A_1$ and $A_2$:
\begin{eqnarray*}
\centering
Y_1 &=& (X_1\times X_2\times X_4)+(A_1\times X_2 \times X_7)+(X_3 \times X_5 \times X_6)+2^{(X_8 \times X_9)}\\
&& +\phantom{1}(A_1\times X_1\times X_5)+(A_2 \times X_3 \times X_5)\\
Y_2 &=& X_1+X_2+(X_3\times X_4 \times X_5)+(A_1 \times X_3)+(A_1 \times A_2 \times X_1).
\end{eqnarray*}

We estimated regressions in three scenarios representing differing types of functional form misspecification: 
\begin{eqnarray*}
\text{Scenario 1:  }Y_1 &=& \beta_1 X_1 + \beta_2 X_2 + \beta_3 X_3 + \beta_4 X_5 + \beta_5 X_6 + \beta_6  X_7 + \beta_7 X_8 + \beta_8 X_9\\
\text{Scenario 2:  } Y_2 &=& \gamma_1 X_1 + \gamma_2 X_2 + \gamma_3 X_3 + \gamma_4 X_4+ \gamma_5 X_5 + \gamma_6 X_6 + \gamma_7  X_7 + \gamma_8 X_8 + \gamma_9 X_9 \\
\text{Scenario 3:  } Y_2 &=& \zeta_1 X_1 +\zeta_2 X_4+ \zeta_3 X_6 +  \zeta_4 X_7 + \zeta_5 X_8 +  \zeta_6 X_9.
\end{eqnarray*}
Complete results for 500 draws from the population with $N=1,000$ and $N=10,000$ are given in Web Tables 1 and 2. Simulation data and complete analytic code to reproduce the simulation analyses are available online: \url{https://github.com/zinka88/Fair-Regression}.

 \begin{table}
\scriptsize
  \caption*{Web Table 1: Simulation Results, N=1,000}
  \label{tab:commands1k}
  \begin{tabular}{clrcrrr}
  &  & & \multicolumn{1}{c}{Predictive} & \multicolumn{2}{c}{Net}  & \multicolumn{1}{c}{}\\ 
   &  & & \multicolumn{1}{c}{Ratio} & \multicolumn{2}{c}{Compensation}  & \multicolumn{1}{c}{Fair}\\ 
\textit{Scenario} &  Method   &  $R^2$   & $g_1$ &  \multicolumn{1}{c}{$g_1$}  & \multicolumn{1}{c}{$g_2$} & \multicolumn{1}{c}{Covariance} \\    
\noalign{\medskip}\hline\noalign{\medskip}
1 & \color{darkgray} Net Compensation, $\lambda=5000$ &	-2901.0 & 5.96 &	3231 &	-273 &	-198.6 \\
&  \color{darkgray}Net Compensation, $\lambda=1000$ &	-106.9 & 1.63 & 408 & -270 & -25.0 \\
& \color{darkgray}Average &	-8.8 &	0.98 & 	-15 &	-270 &	0.9 \\
& \color{darkgray} Covariance, $m=0.2$ &	-8.8 &	0.98 &	-15	& -270 &	0.9 \\
& \color{darkgray} Mean Residual Difference, $\lambda=5000$ & -7.0 & 0.96 &	-28 & -270 & 1.7 \\
& \color{darkgray} Mean Residual Difference, $\lambda=1000$ & -2.0 & 0.89 &	-70	& -270 & 4.3 \\ \vspace{3pt}
& \color{darkgray} Weighted Average, $\alpha=0.2$ &	-1.9 &	0.89 & -71 & -270 &	4.4 \\
& Net Compensation Constraint, $z=0.2$ & 0.1 &	0.86 & -92 & -270 & 5.7 \\
& Weighted Average, $\alpha=0.4$ &	3.5 & 0.80 & -128 &	-269 &	7.9 \\
& Weighted Average, $\alpha=0.6$ & 7.3 & 0.72 &	-185 &	-269 & 11.4 \\
& Mean Residual Difference, $\lambda=100$ &  8.7 &	0.67 &	-215 & 	-269 &	13.2 \\ 
& Net Compensation, $\lambda=100$ & 9.1 & 0.65 & -227 &	-269 &	14.0 \\
& Weighted Average, $\alpha=0.8$ &	9.6 & 0.63 & -241 &	-269 & 	14.9 \\
& Net Compensation Constraint, $\alpha=0.6$ & 9.6 & 0.62 & 	-245 & -269 & 15.1 \\
& Net Compensation Constraint, $z=1$ & 10.4 & 0.54 & -298 &	-269 &	18.3 \\
& OLS &	10.4 & 0.54 & -298 & -269 &	18.4 \\
\noalign{\medskip}
2 &  \color{darkgray} Net Compensation, $\lambda=5000$ & -3436.9 & 2.53 & 217 &	43 & -13.3 \\
& \color{darkgray} Net Compensation, $\lambda=1000$ & -85.7 & 1.07 &	9 &	5 &	-0.6 \\
& \color{darkgray} Average &	 -33.5 &0.99 &	-1 &	3 & 0.1 \\
& \color{darkgray} Covariance, $m=0.2$ &	-33.4 &	0.99 & -1 & 3 & 0.1 \\
& \color{darkgray} Mean Residual Difference, $\lambda=5000$ & -26.5 & 0.98 & -3 & 3 &	0.2 \\
& \color{darkgray} Net Compensation Constraint, $z=0.2$ &	-14.3 & 0.96 & -6 &	3 & 0.4 \\
& \color{darkgray} Mean Residual Difference, $\lambda=1000$ & -5.6 & 0.94 & -8	& 2 & 0.5 \\  \vspace{3pt}
& \color{darkgray} Weighted Average, $\alpha=0.2$ &	 -1.5 &	0.93 &	-10 & 	2 & 	0.6 \\
& Net Compensation Constraint, $\alpha=0.6$ & 17.2 & 0.89 & -16	& 1 & 1.0 \\
& Weighted Average, $\alpha=0.4$ &	23.5 & 	0.87 & 	-18	 & 0 &  1.1 \\
& Net Compensation Constraint, $z=1$ & 39.3 &  0.82 & -25 & -1 & 1.5 \\
& Weighted Average, $\alpha=0.6$ &41.3 & 0.82 & -26 & -1 & 1.6 \\
& Mean Residual Difference, $\lambda=100$ & 45.9 & 0.80 & -29 & -2 & 1.8 \\
& Weighted Average, $\alpha=0.8$ & 52.2 & 0.76 & -34 & -3 & 2.1 \\
& Net Compensation, $\lambda=100$ &	54.4 &	0.74 & -37 & -3 & 2.3 \\
& OLS &	56.0 &	0.70 & -43 & -4 & 2.6 \\

\noalign{\medskip}
3 &  \color{darkgray}Net Compensation, $\lambda=5000$ &  -726.2 & 1.00 & 1 & 44	& 0.0 \\
& \color{darkgray} Average & -582.7 & 0.97 & -5 & 39 & 0.3 \\
& \color{darkgray} Covariance, $m=0.2$ & -582.4 & 0.97 & -5 & 39 & 0.3 \\
& \color{darkgray} Net Compensation Constraint, $z=0.2$ & -472.3 & 0.94 & -9 & 35 &	0.6 \\
& \color{darkgray} Mean Residual Difference, $\lambda=5000$  & -395.8 &	0.91 & -12 & 32 & 0.8 \\  
& \color{darkgray} Weighted Average, $\alpha=0.2$ & -358.0 & 0.90 &  -14 & 31 &	0.9 \\
& \color{darkgray} Net Compensation Constraint, $\alpha=0.6$ & -283.7 & 0.87 &-18 & 27 &	1.1 \\
& \color{darkgray} Weighted Average, $\alpha=0.4$ & -183.0 &	0.83 &	-24 & 22 & 1.5 \\
& \color{darkgray} Net Compensation Constraint, $z=1$ & -138.0 & 0.81 & -27 & 19 & 1.6 \\
& \color{darkgray} Mean Residual Difference, $\lambda=100$ & -121.7 & 0.80 &	-28 & 18 & 1.7 \\ \vspace{3pt}
& \color{darkgray} Weighted Average, $\alpha=0.6$ & -57.7 & 0.77 & -33 &	13 & 2.0 \\
& Net Compensation, $\lambda=1000$ & 12.0 & 0.71 & -42 & 6 & 2.6 \\
& Weighted Average, $\alpha=0.8$ & 18.0 & 0.70 & -43 & 5 & 2.6 \\
& Mean Residual Difference, $\lambda=100$ & 37.5 & 0.66 & -48 &	0 &	2.9 \\
& Net Compensation, $\lambda=100$ & 43.5 & 0.64	& -51 &	-3 & 3.1 \\
& OLS &	44.0 & 0.63 & -52 & -4 & 3.2
  \end{tabular}

  \vspace{3pt}
 
   \scriptsize{\textit{Note:} Measures calculated based on cross-validated predicted values and sorted on net compensation. Estimators with negative $R^2$ values are in shaded text.}
   \end{table}

 \begin{table}
\scriptsize
  \caption*{Web Table 2: Simulation Results, N=10,000}
  \label{tab:commands10k}
  \begin{tabular}{clrcrrr}
  &  & & \multicolumn{1}{c}{Predictive} & \multicolumn{2}{c}{Net}  & \multicolumn{1}{c}{}\\ 
   &  & & \multicolumn{1}{c}{Ratio} & \multicolumn{2}{c}{Compensation}  & \multicolumn{1}{c}{Fair}\\ 
\textit{Scenario} &  Method   &  $R^2$   & $g_1$ &  \multicolumn{1}{c}{$g_1$}  & \multicolumn{1}{c}{$g_2$} & \multicolumn{1}{c}{Covariance} \\    
\noalign{\medskip}\hline\noalign{\medskip}
1 &  \color{darkgray}Net Compensation, $\lambda=5000$  & -15.9 &	1.08 & 	51	& -271	& -3.1 \\ 
&   \color{darkgray}Average &  -8.4 &   1.00 &  -1 &  -271  & 0.1 \\ 
&  \color{darkgray}Covariance, $m=0.2$ &  -8.4 & 1.00   & -1 &  -271 & 0.1 \\  \vspace{3pt}
&   \color{darkgray}Weighted Average, $\alpha=0.2$    & -1.3 & 0.90 & -60 & -271 & 3.7 \\   
&  Net Compensation Constraint, $z=0.2$ & 0.8 & 0.88	& -81 &		-271 &	5.0 \\ 
&  Mean Residual Difference, $\lambda=5000$  & 2.6 &	0.85  & -101 &	-271 & 6.2   \\
&  Weighted Average, $\alpha=0.4$    & 4.2 & 0.82		& -120	&  -271 & 7.3 \\ 
&  Weighted Average, $\alpha=0.6$    & 8.2 & 0.73  & -179 &  -271 & 10.9 \\ 
&  Mean Residual Difference, $\lambda=1000$  &  9.8 &   0.67  &  -213  &-271   &  13.1  \\   
&  Net Compensation, $\lambda=1000$  & 10.2 &	0.65 &	-227  &	-271 &	13.9 \\
& Weighted Average, $\alpha=0.8$ & 10.5 & 0.64  &	-238 & -271 &	14.6 \\
& Net Compensation Constraint, $z=0.6$ & 10.6 &	0.63  &	-241 &	 -271	& 14.8 \\
& Mean Residual Difference, $\lambda=100$ &	11.3 &	0.56  &	-286  & -271 &	17.5 \\
& Net Compensation, $\lambda=100$ & 11.3 &	0.56  & -290 &	-271 & 	17.7 \\
& Net Compensation Constraint, $z=1$ & 11.3 & 0.55 &	-297 &	-271 & 18.2 \\
& OLS & 11.3 & 	0.55  & 	-297  & 	-271 & 18.2  \\
\noalign{\medskip}
2 &  \color{darkgray}Average  & -31.2 & 1.00   & 0   &  3 & 0.0 \\ 
& \color{darkgray}Covariance, $m=.2$ & -31.2 & 1.00    & 0   &  3 & 0.0 \\    \vspace{3pt}
& \color{darkgray}Net Compensation Constraint, $\lambda=0.2$ &	-12.2 &	0.97  &-5 	&3 &0.3 \\  
& Weighted Average, $\alpha=0.2$ &0.4 & 0.94  & -9 &		2 & 0.5 \\
& Mean Residual Difference, $\lambda=5000$ & 12.3 &	0.91  &  -12 &  1 & 0.8 \\
& Net Compensation Constraint, $z=0.6$ & 18.8 & 0.90  &	-15  &	1 &	0.9 \\
& Net Compensation, $\lambda=5000$ & 22.6 &	0.89  &	-16 	& 1 & 	1.0 \\
& Weighted Average, $\alpha=0.4$ & 	25.0 &	0.88  & -17 & 0	& 1.1 \\ 
& Net Compensation Constraint, $z=1$ &	40.6 &	0.83  & -24  &	-1 &1.5 \\
& Weighted Average, $\alpha=0.6$ & 42.6 & 0.82  &-26  & -1 & 1.6 \\
& Mean Residual Difference, $\lambda=1000$ &47.1 &	0.80  & -29  & -2 & 1.8 \\
& Weighted Average, $\alpha=0.8$ & 53.1 & 0.76  &	-34 	& -3 & 2.1 \\ 
& Net Compensation, $\lambda=1000$ &	55.3 &	0.74  & -37  &	-3 & 2.3 \\
& Mean Residual Difference, $\lambda=100$ &	56.4 &	0.72  & -41  &	-4 & 2.5 \\
& Net  Compensation, $\lambda=100$ & 56.6 &	0.71 & -42  &	-4 & 2.6 \\
& OLS &	56.6 &	0.70  &	-43  & -4 &	2.6 \\
\noalign{\medskip}
3 &  \color{darkgray}Average & -637.6 & 1.00 &  -1 &  44&  0.0 \\ 
   & \color{darkgray}Covariance, $m=.2$ & -637.5 & 1.00  &  -1 &   44&  0.0 \\ 
   & \color{darkgray}Net Compensation Constraint, $z=0.2$ &-517.1 & 0.96  &	-5 &  40 & 0.3 \\
   & \color{darkgray}Weighted Average, $\alpha=0.2$ & -392.3 & 0.92 & -11 & 34 &	0.7 \\
   & \color{darkgray}Net Compensation Constraint, $z=0.6$ & -311.2 & 0.89 & -15 & 31 & 0.9 \\
   & \color{darkgray}Weighted Average, $\alpha=0.4$ & -201.5 &	0.85  & -21 &  25 & 1.3 \\
   & \color{darkgray}Net Compensation Constraint, $z=1$ & -152.2 & 0.83  & -25 &		22 & 1.5 \\
   & \color{darkgray}Weighted Average, $\alpha=0.6$ & -65.1 & 0.78  &	-32 & 15 & 2.0 \\  \vspace{3pt}
   & \color{darkgray}Mean Residual Difference, $\lambda=5000$ &	-28.7 &	0.75  & -36 &12 & 2.2 \\
   & Weighted Average, $\alpha=0.8$ & 16.7 & 0.70  & -42 & 5 & 2.6 \\
   & Net Compensation, $\lambda=5000$ &	37.3 & 0.67  & -47 & 1 & 2.9 \\
   & Mean Residual Difference, $\lambda=1000$ &	38.7 & 0.66  & -48  & 	0 &	3.0 \\
   & Net Compensation, $\lambda=1000$ & 43.8 & 0.64  & -52 & -3 & 3.2 \\
   & Mean Residual Difference, $\lambda=100$ & 44.0 & 0.63  &	-52 & -4 & 3.2 \\
   & Net Compensation, $\lambda=100$ & 44.1 & 0.63  &	-53  & -4 &   3.2 \\
   & OLS & 44.1 & 0.63  &  -53 & -4 & 3.2

  \end{tabular}

  \vspace{3pt}
 
   \scriptsize{\textit{Note:} Measures calculated based on cross-validated predicted values and sorted on net compensation. Estimators with negative $R^2$ values are in shaded text.}
   \end{table}

\pagebreak
\section*{Web Appendix C: Hierarchical Condition Category (HCC) Variables}

\scriptsize
\begin{singlespace}
  \begin{tabular}{cl}
 \textbf{HCC} & \textbf{Description} \\ 
 1	& HIV/AIDS \\
2	& Septicemia, Sepsis, Systemic Inflammatory Response Syndrome/Shock \\
6	& Opportunistic Infections \\
8	& Metastatic Cancer and Acute Leukemia \\
9	& Lung and Other Severe Cancers \\
10	& Lymphoma and Other Cancers \\
11	& Colorectal, Bladder, and Other Cancers \\
12	& Breast, Prostate, and Other Cancers and Tumors \\
17	& 	Diabetes with Acute Complications \\
18	& 	Diabetes with Chronic Complications \\ 
19	& 	Diabetes without Complication \\
21	& 	Protein-Calorie Malnutrition \\
22	 & 	Morbid Obesity \\
23	 & Other Significant Endocrine and Metabolic Disorders \\
27	& 	End-Stage Liver Disease \\
28	& 	Cirrhosis of Liver \\
29	& 	Chronic Hepatitis \\
33	& 	Intestinal Obstruction/Perforation \\
34	& 	Chronic Pancreatitis \\
35	& 	Inflammatory Bowel Disease \\
39	& 	Bone/Joint/Muscle Infections/Necrosis \\
40	& 	Rheumatoid Arthritis and Inflammatory Connective Tissue Disease \\
46		& Severe Hematological Disorders \\
47	 	&  Disorders of Immunity \\
48	& 	Coagulation Defects and Other Specified Hematological Disorders \\
54	& 	Drug/Alcohol Psychosis \\
55		& Drug/Alcohol Dependence \\
57	 	& Schizophrenia \\
58	& 	Major Depressive, Bipolar, and Paranoid Disorders \\
72 	& 	Spinal Cord Disorders/Injuries \\
75	 	& Myasthenia Gravis/Myoneural Disorders,  Inflammatory and Toxic Neuropathy \\
77	 	& Multiple Sclerosis \\
78	 	& Parkinson's and Huntington's Diseases \\
79	 	& Seizure Disorders and Convulsions \\
80	& 	Coma, Brain Compression/Anoxic Damage \\
84	& 	Cardio-Respiratory Failure and Shock \\
85	& Congestive Heart Failure \\
86	& 	Acute Myocardial Infarction \\
87		& Unstable Angina and Other Acute Ischemic Heart Disease \\
88	& 	Angina Pectoris \\
96	& Specified Heart Arrhythmias \\
99	& 	Cerebral Hemorrhage \\
100	& 	Ischemic or Unspecified Stroke \\
103	& 	Hemiplegia/Hemiparesis \\
107	& 	Vascular Disease with Complications \\
108	& 	Vascular Disease \\
111	& Chronic Obstructive Pulmonary Disease \\
112	& Fibrosis of Lung and Other Chronic Lung Disorders \\
114	& 	Aspiration and Specified Bacterial Pneumonias \\
122	& 	Proliferative Diabetic Retinopathy and Vitreous Hemorrhage \\
134	& 	Dialysis Status \\
135	& 	Acute Renal Failure \\
136	& 	Chronic Kidney Disease, Stage 5 \\ 
137	& 	Chronic Kidney Disease, Severe (Stage 4) \\
161	& 	Chronic Ulcer of Skin, Except Pressure \\
167	& 	Major Head Injury \\
169	& 	Vertebral Fractures without Spinal Cord Injury \\
170	& 	Hip Fracture/Dislocation \\
173	& 	Traumatic Amputations and Complications \\
176	& 	Complications of Specified Implanted Device or Graft \\
186	& 	Major Organ Transplant or Replacement Status \\
188	& 	Artificial Openings for Feeding or Elimination \\
\end{tabular}
\end{singlespace}

\pagebreak
\normalsize
\section*{Web Appendix D: Simulated Analysis Data}
The IBM MarketScan Research Databases analyzed  in Section 3 of the manuscript cannot be distributed online due to their proprietary nature. They also contain protected patient information. Thus, we created a simulated data set with similar properties using key features and relationships from the original  data for reproducibility analyses of our code.
The simulated analysis data described below and accompanying code to complete the analyses are available  online: \url{https://github.com/zinka88/Fair-Regression}.

First, we simulated demographic variables, female and age, by sampling from a Bernoulli distribution, $\text{female} \sim \text{Bernoulli} (0.52)$, and truncated Normal distribution with lower bound $a$ and upper bound $b$: $\text{age}  \sim \text{Normal} (44, 12, a=21,b=63).$
Next, we generated the 62 binary health variables $\boldsymbol{H}=(H_1, ..., H_T )$, each drawn from a Bernoulli distribution and dependent on the demographic variables female and age with coefficients determined by the relationships in the original data.
To create the indicator for MHSUD, $A$, we generated 15 binary MHSUD CCS variables $\boldsymbol{C}= (C_1, ..., C_{15})$ dependent on age, female, and the top six HCCs correlated with MHSUD in the original data. We defined $A=1$ for all observations with at least one MHSUD CCS; 15.7\% of observations in the simulated analysis data had MHSUD compared to 13.8\% in the original data. 

To generate $Y$, we added random noise to an intermediary outcome $\ddot{Y}$ dependent on the input vector $X = \{\text{female},\text{age},\boldsymbol{H},\boldsymbol{C}\}$. We note that while $\boldsymbol{C}$ was used to generate $\ddot{Y}$, it is not used later in  the estimation steps as this information is not currently included in risk adjustment formulas. $\ddot{Y}$ was determined using a 2-part model. First, to capture the 10.5\% of observations without spending in the original data, we generated whether any spending occurred by creating a binary variable $S$ with $S \sim \text{Bernoulli}(p_{S})$, where $p_{S} = \text{logit}^{-1} [ \boldsymbol{\Omega X}]$ and $\boldsymbol{\Omega}$ is a vector of coefficients based on the original data. Next, for observations with positive spending, we generated the amount of spending that occurred using a log-linear model of spending dependent on $\boldsymbol{X}$ to account for the right-skew of the spending outcome:
$$
\ddot{Y}=
\begin{cases}
    0,& \text{if } S=0\\
    e^{\boldsymbol{\Phi X}}, & \text{if } S=1, \\
\end{cases}
$$
\noindent where $\boldsymbol{\Phi}$ is a vector of coefficients based on the original data. Lastly, we sampled from a truncated normal centered around each observation in $\ddot{Y}$ to add noise to the generated outcome: $Y_{k}  \sim \text{Normal}(\ddot{Y_k}, 6000, a=0, b=\text{max}(Y)),$ where $Y_k$ is the predicted outcome for observation $k$ in the simulated data. The final simulated spending outcome ranged from \$0 to \$297,206 with a mean of \$5,817 and median of \$4,881. The average spending for enrollees with MHSUD was \$6,812 versus \$5,632 for enrollees without MHSUD. $R^2$ under OLS was  19.7\%. 

The results from the simulated analysis data are shown in Web Table 3. As demonstrated in our data analysis presented in Section 3 of the main text, we likewise find that the constrained and penalized estimation methods improve fairness measures without a significant decrease in $R^2$. The relative rankings of all the methods are similar, although we  highlight that  net compensation penalized regression performs even more similarly to average constrained  and covariance constrained regression methods here.\\

\begin{table}
\small
  \caption*{Web Table 3: Performance of Constrained and Penalized Regression Methods in Simulated  Data}
  \label{tab:commandsdata}
  \begin{tabular}{rcccrrrr}
     & & \multicolumn{2}{c}{Predictive} & \multicolumn{2}{c}{Net} & \multicolumn{1}{c}{Mean}  & \multicolumn{1}{c}{}\\ 
     & & \multicolumn{2}{c}{Ratio} & \multicolumn{2}{c}{Compensation} &  \multicolumn{1}{c}{Residual} & \multicolumn{1}{c}{Fair}\\ 
  \textbf{Method}    &  $R^2$   & $g$ & $g^c$ & \multicolumn{1}{c}{$g$} & \multicolumn{1}{c}{$g^c$} & Difference & Covariance \\    
\noalign{\medskip}\hline\noalign{\medskip}
    Net Compensation$^\dagger$  & 18.5\% &   1.001 & 1.000 & \$6 & -\$1 &  \$7 & -1\\
 \\
    Average  & 18.6 &   0.999 & 1.000 &-5 & 1 & -7 & 1\\ 
    \\
    Covariance & 18.6 &  0.999 & 1.000 & -5 & 1 &  -7 & 1\\
    \\
             Weighted Average$^\mp$    & 19.0 &  0.984 & 1.004 & -106 & 20 & -127 & 17 \\ 
          \\
             Mean Residual Difference$^\oplus$  & 19.6 & 0.947 & 1.012 & -364 & 68 & -432 & 57\\
    \\
        OLS & 19.7 & 0.925 & 1.017 & -512 & 95 & -607 & 80 \\ 
\noalign{\medskip}\hline
  \end{tabular}
  
  \vspace{6pt}
 
   \scriptsize{$^\dagger \lambda=20000$, $^\mp \alpha=0.2$, $^\oplus \lambda=30000$\\
   \textit{Note:} Measures calculated based on cross-validated predicted values and sorted on net compensation. Best performing hyperparameters for each estimator (with respect to fairness measures) are displayed. Performance for covariance method was the same for all $m$. $g^c$ is the complement of g.}
   \end{table}

\normalsize

  \bibliographystyle{asa} 
 \bibliography{sample-bibliography}

\begin{thebibliography}{48}
\newcommand{\enquote}[1]{``#1''}
\expandafter\ifx\csname natexlab\endcsname\relax\def\natexlab#1{#1}\fi

\bibitem[{Ash and Ellis(2012)}]{AshEllis12}
Ash, {\relax AS}. and Ellis, {\relax RP}. (2012), \enquote{Risk-adjusted
  payment and performance assessment for primary care.} \textit{Med Care}, 50,
  643--653.

\bibitem[{Bansal et~al.(2014)Bansal, Sinha, and Zhao}]{Bansal14}
Bansal, G., Sinha, A., and Zhao, H. (2014), \enquote{Tuning Data Mining Methods
  for Cost-Sensitive Regression: A Study in Loan Charge-Off Forecasting,}
  \textit{Journal of Management Information Systems}, 25, 315--336.

\bibitem[{Bechavod and Ligett(2018)}]{Bechavod17}
Bechavod, Y. and Ligett, K. (2018), \enquote{Penalizing Unfairness in Binary
  Classification,} arXiv pre-print. \url{arxiv.org/abs/1707.00044}.

\bibitem[{Bergquist et~al.(2019)Bergquist, Layton, McGuire, and
  Rose}]{Bergquist19}
Bergquist, {\relax SL}., Layton, {\relax TJ}., McGuire, {\relax TG}., and Rose,
  S. (2019), \enquote{Data transformations to improve the performance of health
  plan payment methods,} \textit{Journal of Health Economics}, 66, 195 -- 207.

\bibitem[{Berk et~al.(2017{\natexlab{a}})Berk, Heidari, Jabbari, Joseph,
  Kearns, Morgenstern, Neel, and Roth}]{Berk17b}
Berk, R., Heidari, H., Jabbari, S., Joseph, M., Kearns, M., Morgenstern, J.,
  Neel, S., and Roth, A. (2017{\natexlab{a}}), \enquote{A Convex Framework for
  Fair Regression,} arXiv pre-print. \url{arxiv.org/abs/1706.02409}.

\bibitem[{Berk et~al.(2017{\natexlab{b}})Berk, Heidari, Jabbari, Kearns, and
  Roth}]{Berk17}
Berk, R., Heidari, H., Jabbari, S., Kearns, M., and Roth, A.
  (2017{\natexlab{b}}), \enquote{Fairness in Criminal Justice Risk Assessments:
  The State of the Art,} arXiv pre-print. \url{arxiv.org/abs/1703.09207}.

\bibitem[{Calders et~al.(2013)Calders, Karim, Kamiran, Ali, and
  Zhang}]{Calders13}
Calders, T., Karim, A., Kamiran, F., Ali, W., and Zhang, X. (2013),
  \enquote{Controlling Attribute Effect in Linear Regression,} in \textit{2013
  IEEE 13th International Conference on Data Mining}, {Dallas, TX, USA}.

\bibitem[{Calmon et~al.(2017)Calmon, Wei, Ramamurthy, and Varshney}]{Calmon17}
Calmon, {\relax FP}., Wei, D., Ramamurthy, {\relax KN}., and Varshney, {\relax
  KR}. (2017), \enquote{Optimized Data Pre-Processing for Discrimination
  Prevention,} arXiv pre-print. \url{arxiv.org/abs/1704.03354}.

\bibitem[{Carey(2017)}]{Carey17}
Carey, C. (2017), \enquote{Technological Change and Risk Adjustment: Benefit
  Design Incentives in {Medicare Part D},} \textit{American Economic Journal:
  Economic Policy}, 9, 38--73.

\bibitem[{Chouldechova(2017)}]{Chouldechova17}
Chouldechova, A. (2017), \enquote{Fair prediction with disparate impact: A
  study of bias in recidivism prediction instruments,} arXiv pre-print.
  \url{arxiv.org/abs/1610.07524}.

\bibitem[{Chouldechova and Roth(2018)}]{Chouldechova18}
Chouldechova, A. and Roth, A. (2018), \enquote{The Frontiers of Fairness in
  Machine Learning,} arXiv pre-print. \url{arxiv.org/abs/1810.08810}.

\bibitem[{Corbett-Davies and Goel(2018)}]{Davies18}
Corbett-Davies, S. and Goel, S. (2018), \enquote{The Measure and Mismeasure of
  Fairness: A Critical Review of Fair Machine Learning,} arXiv pre-print.
  \url{arxiv.org/abs/1808.00023}.

\bibitem[{Dwork et~al.(2018)Dwork, Immorlica, Kalai, and Leiserson}]{Dwork17}
Dwork, C., Immorlica, N., Kalai, {\relax AT}., and Leiserson, M. (2018),
  \enquote{Decoupled classifiers for fair and efficient machine learning,}
  \textit{Proceedings of Machine Learning Research}, 81, 119--133.

\bibitem[{Ellis et~al.(2018)Ellis, Martins, and Rose}]{Ellis17}
Ellis, {\relax RP}., Martins, B., and Rose, S. (2018), \enquote{Risk Adjustment
  for Health Plan Payment,} in \textit{Risk Adjustment, Risk Sharing and
  Premium Regulation in Health Insurance Markets: Theory and Practice}, {edited
  by TG. McGuire, and RC. van Kleef. Amsterdam: Elsevier}.

\bibitem[{{\relax El Mhamdi} et~al.(2018){\relax El Mhamdi}, Guerraoui, Hoang,
  and Maurer}]{Mhamdi18}
{\relax El Mhamdi}, {\relax EM}., Guerraoui, R., Hoang, {\relax LN}., and
  Maurer, A. (2018), \enquote{Removing Algorithmic Discrimination (With Minimal
  Individual Error),} arXiv pre-print. \url{arxiv.org/abs/1806.02510}.

\bibitem[{Ericson et~al.(2017)Ericson, Geissler, and Lubin}]{Geissler17}
Ericson, {\relax KM}., Geissler, K., and Lubin, B. (2017), \enquote{The Impact
  of Partial-Year Enrollment on the Accuracy of Risk Adjustment Systems: A
  Framework and Evidence,} NBER Working Paper \#23765.
  \url{nber.org/papers/w23765}.

\bibitem[{Fu et~al.(2018)Fu, Narasimhan, and Boyd}]{cvxr}
Fu, A., Narasimhan, B., and Boyd, S. (2018), \enquote{{CVXR:} An R Package for
  Discplined Convex Optimization,}
  \url{web.stanford.edu/~boyd/papers/pdf/cvxr_paper.pdf}, online; 30 July 2018.

\bibitem[{Geruso et~al.(2017)Geruso, Layton, and Prinz}]{Geruso17}
Geruso, M., Layton, {\relax TJ}., and Prinz, D. (2017), \enquote{Screening in
  Contract Design: Evidence from the {ACA} Health insurance exchanges,} NBER
  Working Paper \#22832. \url{nber.org/papers/w22832}.

\bibitem[{Gibney(2018)}]{nature}
Gibney, E. (2018), \enquote{The ethics of computer science: this researcher has
  a controversial proposal,} \url{nature.com/articles/d41586-018-05791-w},
  online; 9 August 2018.

\bibitem[{Hardt et~al.(2016)Hardt, Price, and Srebro}]{Hardt16}
Hardt, M., Price, E., and Srebro, N. (2016), \enquote{Equality of Opportunity
  in Supervised Learning,} arXiv pre-print. \url{arxiv.org/abs/1610.02413}.

\bibitem[{Hastie et~al.(2009)Hastie, Tibshirani, and Friedman}]{statlearn}
Hastie, T., Tibshirani, R., and Friedman, J. (2009), \textit{The Elements of
  Statistical Learning: Data Mining, Inference, and Prediction. 2nd Edition},
  {New York City: Springer}.

\bibitem[{Jacobs and Sommers(2015)}]{Jacobs15}
Jacobs, {\relax DB}. and Sommers, {\relax BD}. (2015), \enquote{Using Drugs to
  Discriminate -- Adverse Selection in the Insurance Marketplace,}
  \textit{NEJM}, 372, 399--402.

\bibitem[{Johndrow and Lum(2017)}]{Johndrow17}
Johndrow, {\relax JE}. and Lum, K. (2017), \enquote{An algorithm for removing
  sensitive information: application to race-independent recidivism
  prediction,} arXiv pre-print. \url{arxiv.org/abs/1703.04957}.

\bibitem[{Kamiran and Calders(2009)}]{Kamiran12}
Kamiran, F. and Calders, T. (2009), \enquote{Classifying without
  Discrimination,} in \textit{2009 2nd International Conference on Computer,
  Control and Communication}, {Karachi, Pakistan}.

\bibitem[{Kamishima et~al.(2012)Kamishima, Akaho, Asoh, and
  Sakuma}]{Kamishima12}
Kamishima, T., Akaho, S., Asoh, H., and Sakuma, J. (2012),
  \enquote{Fairness-Aware Classifier with Prejudice Remover Regularizer,} in
  \textit{Machine Learning and Knowledge Discovery in Databases}, vol. 7524.

\bibitem[{Kautter et~al.(2014)Kautter, Pope, Ingber, Freeman, Patterson, Cohen,
  and Keenan}]{Kautter14}
Kautter, J., Pope, {\relax GC}., Ingber, M., Freeman, S., Patterson, L., Cohen,
  M., and Keenan, P. (2014), \enquote{The {HHS-HCC} Risk Adjustment Model for
  Individual and Small Group Markets under the {Affordable Care Act},}
  \textit{Medicare \& Medicaid Research Review}, 4, mmrr2014--004--03--a03.

\bibitem[{Kleinberg et~al.(2018)Kleinberg, Ludwig, Mullainathan, and
  Rambachan}]{Kleinberg18}
Kleinberg, J., Ludwig, J., Mullainathan, S., and Rambachan, A. (2018),
  \enquote{Algorithmic Fairness,} \textit{AEA Papers and Proceedings}, 108,
  22--27.

\bibitem[{Kleinberg et~al.(2016)Kleinberg, Mullainathan, and
  Raghavan}]{Kleinberg16}
Kleinberg, J., Mullainathan, S., and Raghavan, M. (2016), \enquote{Inherent
  Trade-Offs in the Fair Determination of Risk Scores,} arXiv pre-print.
  \url{arxiv.org/abs/1609.05807}.

\bibitem[{Layton et~al.(2017)Layton, Ellis, McGuire, and {\relax van
  Kleef}}]{Layton17}
Layton, {\relax TJ}., Ellis, {\relax RP}., McGuire, {\relax TG}., and {\relax
  van Kleef}, {\relax RC}. (2017), \enquote{Measuring efficiency of health plan
  payment systems in managed competition health insurance markets,}
  \textit{Journal of Health Economics}, 56, 237--255.

\bibitem[{{McGuire} and {van Kleef}(2018)}]{rabook}
{McGuire}, T. and {van Kleef}, R. (eds.) (2018), \textit{Risk Adjustment, Risk
  Sharing and Premium Regulation in Health Insurance Markets}, {Amsterdam:
  Elsevier}.

\bibitem[{McGuire et~al.(2013)McGuire, Glazer, Newhouse, Normand, Shi, Sinaiko,
  and Zuvekas}]{McGuire14a}
McGuire, {\relax TG}., Glazer, J., Newhouse, {\relax JP}., Normand, {\relax
  SL}., Shi, J., Sinaiko, {\relax AD}., and Zuvekas, {\relax SH}. (2013),
  \enquote{Integrating Risk Adjustment and Enrollee Premiums in Health Plan
  Payment,} \textit{Journal of Health Economics}, 32, 1263--1277.

\bibitem[{Montz et~al.(2016)Montz, Layton, Busch, Ellis, Rose, and
  McGuire}]{Montz16}
Montz, E., Layton, {\relax TJ}., Busch, {\relax AB}., Ellis, {\relax RP}.,
  Rose, S., and McGuire, {\relax TG}. (2016), \enquote{Risk adjustment
  simulation: Plans may have incentives to distort mental health and substance
  use coverage,} \textit{Health Affairs}, 35, 1022--1028.

\bibitem[{O'Neil(2017)}]{nyt}
O'Neil, C. (2017), \textit{Weapons of Math Destruction}, {New York City:
  Broadway Books}.

\bibitem[{Park and Basu(2018)}]{Park18}
Park, S. and Basu, A. (2018), \enquote{Alternative evaluation metrics for risk
  adjustment methods,} \textit{Health Economics}, 27, 984--1010.

\bibitem[{Pope et~al.(2004)Pope, Kautter, Ellis, et~al.}]{Pope04}
Pope, {\relax GC}., Kautter, J., Ellis, {\relax RP}., et~al. (2004),
  \enquote{Risk Adjustment for Medicare Capitation Payments Using the {CMS-HCC}
  Model,} \textit{Health Care Financing Review}, 25, 119--141.

\bibitem[{Rose(2016)}]{Rose16}
Rose, S. (2016), \enquote{A Machine Learning Framework for Plan Payment Risk
  Adjustment,} \textit{Health Services Research}, 51, 2358--2374.

\bibitem[{Rose et~al.(2017)Rose, Bergquist, and Layton}]{Rose17}
Rose, S., Bergquist, {\relax SL}., and Layton, {\relax TJ}. (2017),
  \enquote{Computational health economics for identification of unprofitable
  health care enrollees,} \textit{Biostatistics}, 18, 682--694.

\bibitem[{Rose and McGuire(2019)}]{RoseMcguire19}
Rose, S. and McGuire, {\relax TG}. (2019), \enquote{{Limitations of
  {P}-{Values} and {R}-squared for {Stepwise} {Regression} {Building}: {A}
  {Fairness} {Demonstration} in {Health} {Policy} {Risk} {Adjustment}},}
  \textit{The American Statistician}, 73, 152--156.

\bibitem[{Shepard(2016)}]{Shepard16}
Shepard, M. (2016), \enquote{Hospital Network Competition and Adverse
  Selection: Evidence from the {Massachusetts Health Insurance Exchange},} NBER
  Working Paper \#22600. \url{nber.org/papers/w22600}.

\bibitem[{Shrestha et~al.(2018)Shrestha, Bergquist, Montz, and
  Rose}]{Shrestha17}
Shrestha, A., Bergquist, {\relax SL}., Montz, E., and Rose, S. (2018),
  \enquote{Mental Health Risk Adjustment with Clinical Categories and Machine
  Learning,} \textit{Health Services Research}, 53, 3189--3206.

\bibitem[{{\relax van Kleef} et~al.(2017){\relax van Kleef}, McGuire, {\relax
  van Vliet}, and {\relax van de Ven}}]{VanKleef17b}
{\relax van Kleef}, {\relax RC}., McGuire, {\relax TG}., {\relax van Vliet},
  R., and {\relax van de Ven}, W. (2017), \enquote{Improving risk equalization
  with constrained regression,} \textit{The European Journal of Health
  Economics}, 18, 1137--1156.

\bibitem[{{\relax van Kleef} et~al.(2013){\relax van Kleef}, {\relax van
  Vliet}, and {\relax Van de Ven}}]{VanKleef13}
{\relax van Kleef}, {\relax RC}., {\relax van Vliet}, {\relax RC}., and {\relax
  Van de Ven}, {\relax WP}. (2013), \enquote{Risk equalization in {The
  Netherlands}: an empirical evaluation,} \textit{Expert Rev Pharmacoecon
  Outcomes Res}, 13, 829--839.

\bibitem[{{\relax van Kleef} et~al.(2018){\relax van Kleef}, {\relax van
  Vliet}, and {\relax van de Ven}}]{VanKleef18c}
{\relax van Kleef}, {\relax RC}., {\relax van Vliet}, {\relax RC}., and {\relax
  van de Ven}, {\relax WP}. (2018), \enquote{Health plan payment in the
  {Netherlands},} in \textit{Risk Adjustment, Risk Sharing and Premium
  Regulation in Health Insurance Markets: Theory and Practice}, {edited by TG.
  McGuire, and RC. van Kleef. Amsterdam: Elsevier}.

\bibitem[{Withagen-Koster et~al.(2018)Withagen-Koster, {\relax van Kleef}, and
  Eijkenaar}]{VanKleef18}
Withagen-Koster, {\relax AA}., {\relax van Kleef}, {\relax RC}., and Eijkenaar,
  F. (2018), \enquote{Examining unpriced risk heterogeneity in the Dutch health
  insurance market,} \textit{The European Journal of Health Economics}, 19,
  1351--1363.

\bibitem[{Zafar et~al.(2017{\natexlab{a}})Zafar, Valera, Rodriguez, and
  Gummadi}]{Zafar17b}
Zafar, {\relax MB}., Valera, I., Rodriguez, {\relax MG}., and Gummadi, {\relax
  KP}. (2017{\natexlab{a}}), \enquote{Fairness Beyond Disparate Treatment \&
  Disparate Impact: Learning Classification without Disparate Mistreatment,}
  arXiv pre-print. \url{arxiv.org/abs/1610.08452}.

\bibitem[{Zafar et~al.(2017{\natexlab{b}})Zafar, Valera, Rodriguez, and
  Gummadi}]{Zafar17}
--- (2017{\natexlab{b}}), \enquote{Fairness Constraints: Mechanisms for Fair
  Classification,} arXiv pre-print. \url{arxiv.org/abs/1507.05259}.

\bibitem[{Zemel et~al.(2013)Zemel, We, Swersky, Pitassi, and Dwork}]{Zemel13}
Zemel, R., We, Y., Swersky, K., Pitassi, T., and Dwork, C. (2013),
  \enquote{Learning Fair Representations,} in \textit{Proceedings of the 30th
  International Conference on Machine Learning, PMLR}, vol.~28, pp. 325--333,
  {Atlanta, Georgia, USA}.

\bibitem[{Zliobaite et~al.(2011)Zliobaite, Kamiran, and Calders}]{Zliobaite11}
Zliobaite, I., Kamiran, F., and Calders, T. (2011), \enquote{Handling
  Conditional Discrimination,} in \textit{2011 IEEE 11th International
  Conference on Data Mining}, pp. 992--1001, {Vancouver, BC, Canada}.

\end{thebibliography}

\end{document}